\documentclass[
    ,final            
  ]
  {aipproc}
\usepackage{epstopdf}
\layoutstyle{6x9}

\begin{document}

\title{Electroweak Physics at Jefferson Lab}

\classification{11.30.Er, 12.15.Mm, 13.66.Lm, 13.88.+e,
95.30.Cq, 14.70.Pw, 25.30.Bf, 25.30.Rw, 95.35.+d}
\keywords      {Parity violation, fundamental symmetries, neutral currents, gauge boson, dark matter}

\author{R.~D.~McKeown}{
address ={Thomas Jefferson National Accelerator Facility, Newport News, Virginia 23606, USA},
altaddress={Department of Physics, College of William and Mary, Williamsburg, Virginia 23187, USA}}

\begin{abstract}
The Continuous Electron Beam Accelerator Facility (CEBAF) at the
Thomas Jefferson National Accelerator Facility provides CW electron
beams with high intensity, remarkable stability, and a high degree of polarization.
These capabilities offer new and unique opportunities to
search for novel particles and forces that would require extension of
the standard model. CEBAF is presently undergoing an
upgrade that includes doubling the energy of the electron
beam to 12~GeV and enhancements to the experimental equipment. This
upgraded facility will provide increased capability to address new
physics beyond the standard model.
\end{abstract}

\maketitle

\section{Introduction}

Since 1995, the CEBAF facility at Jefferson Laboratory has operated
high-duty factor (continuous) beams of electrons incident on three
experimental halls (denoted A, B, and C), each with a unique set of experimental
equipment. As a result of advances in the performance of novel
superconducting radiofrequency (SRF) accelerator technology, the
electron beam has exceeded the original 4~GeV energy specification,
and beams with energies up to 6~GeV with currents up to 100~$\mu$A
have been delivered for the experimental program. In addition, the
development of advanced GaAs photoemission sources has enabled high
quality polarized beam with polarizations up to 85\%. The facility
serves an international scientific user community of over 1200
scientists, and to date over 160 experiments have been completed.

The physics research program at CEBAF has generally focused on the structure of
hadronic systems and their excitations, including mesons, nucleons, and atomic nuclei.
This research program has motivated many technical developments such as high-precision polarimeters,
beam monitoring and feedback systems, and specialized magnetic spectrometers and detector systems.
Progress in the measurement of parity-violating asymmetries at the part-per-million level has been particularly impressive, and this process
has been utilized to provide stringent constraints on the strange quark content of the nucleon.

Recently, it has been recognized that the unique capabilities of CEBAF can be employed to perform precision tests of the standard model that may
reveal the existence of new particles and/or forces. In addition, high intensity beams in the 0.1-10~GeV range can be used to search for new vector bosons in
this mass range. As a result, an exciting new research direction for the Laboratory is now being developed to probe unknown physics beyond the standard model.

\section{Physics Beyond the Standard Model}

The discoveries of dark matter, dark energy, and the flavor
oscillations of neutrinos (associated with their small but finite
masses) are all indications that the standard model of the strong
and electroweak interactions requires modification. In addition,
there are theoretical motivations for extending the standard model
associated with protecting the Higgs mass from uncontrolled loop
corrections. These issues generally lead to the view that the
standard model is part of a larger theoretical framework, and such
an extension of the theory should lead to observational
consequences.

There are many experimental activities throughout the world focused on
discovering new physics associated with particles and forces that are not
included in the standard model. These activities generally fall into one
of two general approaches. One is to advance the high energy
frontier with higher energy particle accelerators in an attempt to
obtain evidence for new particles that can be produced at higher
energy. This is the approach that has recently been employed at the Tevatron at Fermilab
and will in the future be the primary focus of the Large Hadron Collider at CERN.
The other method is to perform high precision measurements
at lower energies to provide information that can indicate the
properties of a more complete theory. Indeed, such precision
measurements yield complementary information to that which can be
obtained at the energy frontier. For a review of this approach,
including discussion of parity-violating electron scattering, see
\cite{MJRM}.

\section{Parity-violating Electron Scattering}

 CEBAF is an ideal facility for studies of
parity violating electron scattering. The neutral weak interaction
violates parity symmetry, and so measurement of parity violation in
electron scattering is a consequence of interference between neutral weak
and electromagnetic amplitudes. Extensions of the standard model may
involve the existence of new massive ($M> M_Z$) neutral bosons that
couple to electrons and/or quarks. Such phenomena, or other similar
processes, would lead to changes in the parity-violating interaction
with electrons and thus manifest themselves in parity violation
experiments.

\begin{figure}
\includegraphics[width=4.5 in]{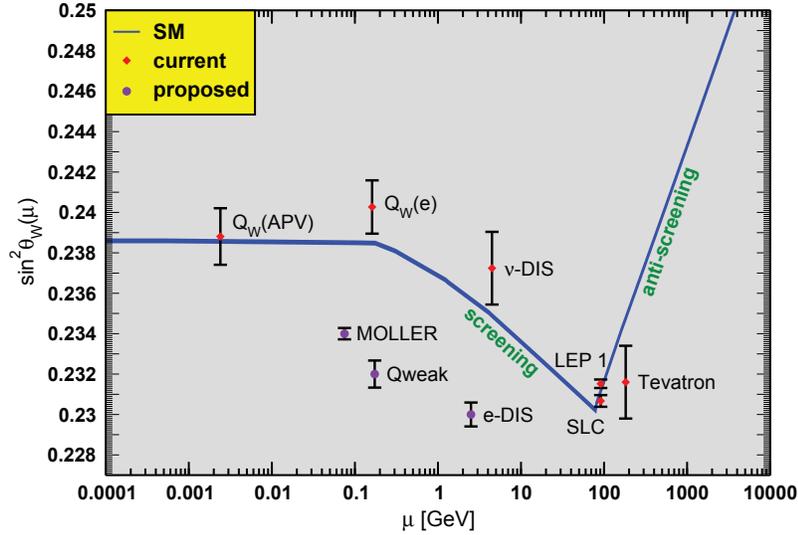}
\caption{\label{fig:PV} The standard model prediction for $\sin^2 \theta_W$, along with previous measurements and expected results for future experiments at Jefferson Lab \cite{erler}.}
\end{figure}

The usual method for experimental studies of parity-violating electron scattering involves a longitudinally polarized electron beam incident on an unpolarized target. One can periodically flip the helicity of the electron beam and measure the fractional change in the scattering rate, known as the parity-violating asymmetry $A_{PV}$:
 \begin{equation}
A_{PV} = {d\sigma_R - d\sigma_L \over d\sigma_R + d\sigma_L}
\end{equation}
 where $d\sigma_L$ and $d\sigma_R$ are the differential cross sections for left and right helicity electrons, respectively.
This asymmetry is due to interference of the neutral weak and electromagnetic interactions, and for electron-hadron interactions can generally be written
\begin{equation}
    A_{PV} \simeq \left[- G_F Q^2 \over 4 \sqrt{2} \pi \alpha \right]\times {\cal F}
\end{equation}
where $\cal F$ depends on the detailed properties of the target and the weak interaction and may contain information related to new physics.
For low $Q^2$ one generally finds the asymmetry to be quite small, parts per million, and so precise measurements are particularly challenging. A fundamental experimental issue is to control the beam properties under helicity reversal in order to reduce systematic errors associated with beam changes (such as slight position or angle shifts).

The strength of the neutral weak interaction is parameterized in the standard model by the weak mixing angle $\theta_W$. This parameter is very precisely determined at the $Z$ boson mass by $e^+$-$e^-$ collider experiments. The two best measurements (which differ by more than $2\sigma$) have uncertainties of 0.00029 and 0.00026, and can be combined to yield the average value $\sin^2 \theta_W =  0.23116  \pm 0.00013 $ \cite{pdg}. Radiative corrections associated with standard model physics predicts a ``running'' of this coupling to $\sin^2 \theta_W =  0.2388$ at $Q^2 =0$. Additional particles at high mass (larger than $M_Z$) would generally modify these radiative corrections, leading to a different value of $\sin^2 \theta_W $ at $Q^2=0$. Thus precise measurements of the neutral weak interaction at low $Q^2 \ll M_Z^2$ can reveal the presence of particles and forces not present in the standard model.

A major new experiment to study parity-violation in elastic electron-proton scattering at low $Q^2$ is presently underway in Hall C at Jefferson Lab\cite{Qweak}.
This experiment, known as $Q_{weak}$, will be taking data during 2011 and 2012. In addition,
there are presently 2 new proposals to perform parity violation measurements at the
upgraded CEBAF. One would use a solenoidal magnetic
spectrometer system (SOLID) to study parity-violating deep inelastic
scattering \cite{SOLID}. The other proposal involves the
construction of a novel dedicated toroidal spectrometer to study
parity-violating M{\o}ller scattering \cite{Moller}. Both
experiments will require construction of substantial new
experimental equipment (beyond the scope of the present upgrade
project) and propose to be sited in experimental Hall A.

 Fig.~\ref{fig:PV} shows the standard model prediction (in "minimal subtraction", or $\overline{\rm MS}$, renormalization scheme) for $\sin^2 \theta_W$ as a function of energy scale $\mu$. The value at the mass of the $Z$ boson is fit to the $e^+-e^-$ data. Also shown are results from atomic parity violation, parity violating M{\o}ller scattering at SLAC (E158), and results from deep inelastic neutrino scattering.  (It should be noted that the nuclear corrections for the deep inelastic neutrino scattering results are still a subject of substantial discussion.) The projected results
for $\sin^2 \theta_W$ for the future Jefferson Lab experiments are shown at the correct energy scale, but arbitrary values of $\sin^2 \theta_W$, to illustrate the expected experimental uncertainty.

\subsection{Parity Violation in Elastic Electron Scattering}

Studies of the parity-violating asymmetry in elastic electron scattering have been utilized for many years to elucidate the role of strange quark-antiquark pairs in the structure of the nucleon \cite{Annrev00}. In addition to providing technical advances that are necessary for future high-precision measurements, these experiments have demonstrated that the strange quark electric form factor $G_E^s (Q^2)$ is constrained to be very small at low $Q^2$ (note that $G_E^s(Q^2=0)=0$ since there is no net strangeness in the nucleon). A recent global analysis \cite{Liu} of the low $Q^2$ data yielded the result $G_E^s [Q^2 = 0.1~(\rm{GeV}/c)^2] =  -0.008 \pm 0.016$. Thus by performing a measurement at lower $Q^2=0.026 ~(\rm{GeV}/c)^2$ the $Q_{weak}$ experiment will reduce the effect of strange quarks, as well as reduce the uncertainty due to electroweak radiative corrections associated with $\gamma-Z$ ``box'' diagrams.

The $Q_{weak}$ experiment will utilize a $150~\mu$A beam of 85\% polarized 1.2~GeV electrons incident on a 35~cm long liquid hydrogen target. The scattered electrons are momentum analyzed by a toroidal magnet and detected in quartz bar Cerenkov detectors. Custom electronics is used to integrate the charge from the Cerenkov detector photomultiplier tubes for each helicity state. The helicity state of the beam is changed at a rate of up to 960/sec. The parity-violating asymmetry measurement precision is projected to be 2.5\%, resulting in a measurement of the weak charge $Q_w^p$ of 4.1\%, and an uncertainty in $\sin^2 \theta_W$ of 0.0008. Such a measurement is sensitive to new neutral weak $Z^\prime$ particles in the $0.5-1$~TeV mass range and to leptoquarks up to a mass of about 6~TeV.

\subsection{Parity Violation in Deep Inelastic Scattering}

Parity violation in deep inelastic scattering was in fact the first experimental demonstration that the neutral current violated parity as expected in the standard model \cite{Prescott}. Such measurements probe the parity violating electron-quark interaction, which is calculable in the standard model. Corrections to the parity-violating asymmetry arise from issues related to the quark distribution functions in the nucleon due to strange quarks and to charge symmetry violation. In addition, there are potential higher twist corrections.

An initial experiment was run with 6~GeV beam in 2009, and is expected to yield new information on the axial quark couplings. A new experiment, known as SOLID \cite{SOLID}, has been proposed for the 12~GeV program at Jefferson Lab. This new experiment would make measurements over a broad kinematic range in order to constrain the uncertainties associated with charge symmetry violation and higher twist effects. The use of a deuterium target minimizes the uncertainties due to the quark distribution functions. The experimental design involves a solenoidal spectrometer with collimators to select kinematics of the scattered electrons in the range $22^\circ < \theta < 35^\circ$, $x_{Bj} > 0.55$, and $W>2$~GeV. High rate GEM detectors will track the scattered electrons, and a threshold gas Cerenkov detector plus electromagnetic calorimeter will enable pion rejection. Several superconducting solenoids from other previous experiments are being considered for the magnet requirement.

\subsection{Parity Violation in M{\o}ller Scattering}
M{\o}ller scattering is a particularly clean process for high precision tests of the standard model. This is a purely leptonic process, and there are no uncertainties associated with the target structure since the electrons in a hydrogen target are essentially non-interacting and at rest. A previous experiment \cite{E158}, E158, at SLAC yielded an impressive result for the parity-violating asymmetry, $A_{PV} = (-131 \pm 14 (\rm{stat.}) \pm 11 (\rm{syst.})) \times 10^{-9}$, and provided a precise determination of the weak mixing angle at low $Q^2$: $\sin^2 \theta_W = 0.2406 \pm 0.0011 ({\rm stat.}) \pm 0.0009 ({\rm syst.})$.

The proposed Jlab experiment, MOLLER, would utilize a $75 ~\mu$A beam of 11~GeV electrons incident on a liquid hydrogen target. A normally conducting toroidal magnetic spectrometer selects the M{\o}ller scattered electrons in the angular range 5-18~mr. The predicted parity-violating asymmetry is $A_{PV} = 35.6$~ppb, and can be measured with a projected statistical uncertainty of 0.73 ppb in 38 weeks of running. This would provide a substantially improved determination of $\sin^2 \theta_W$ with a statistical uncertainty of 0.00026. It is noteworthy that such an impressive result would be of comparable precision to the best measurements at the $Z$ pole, and so offers the best opportunity to reveal new physics from such measurements.

\section{New Gauge Boson Search}

It is now well established that dark matter comprises 21\% of the energy density of the universe \cite{pdg}. The nature of this dark matter is not presently understood, but there is strong evidence that new particles beyond those included in the standard model are required to explain the dark matter. The most favored scheme involves weakly interacting massive particles, or WIMPs, that could have masses in the TeV range. Such particles may be observable at the Large Hadron Collider, but also could be detected by their weak interactions with ordinary baryonic matter in very sensitive particle detector experiments. In fact, there is a very active program of experiments to search for these cosmological WIMPS at underground laboratories around the world.

\begin{figure}
\includegraphics[width=3. in]{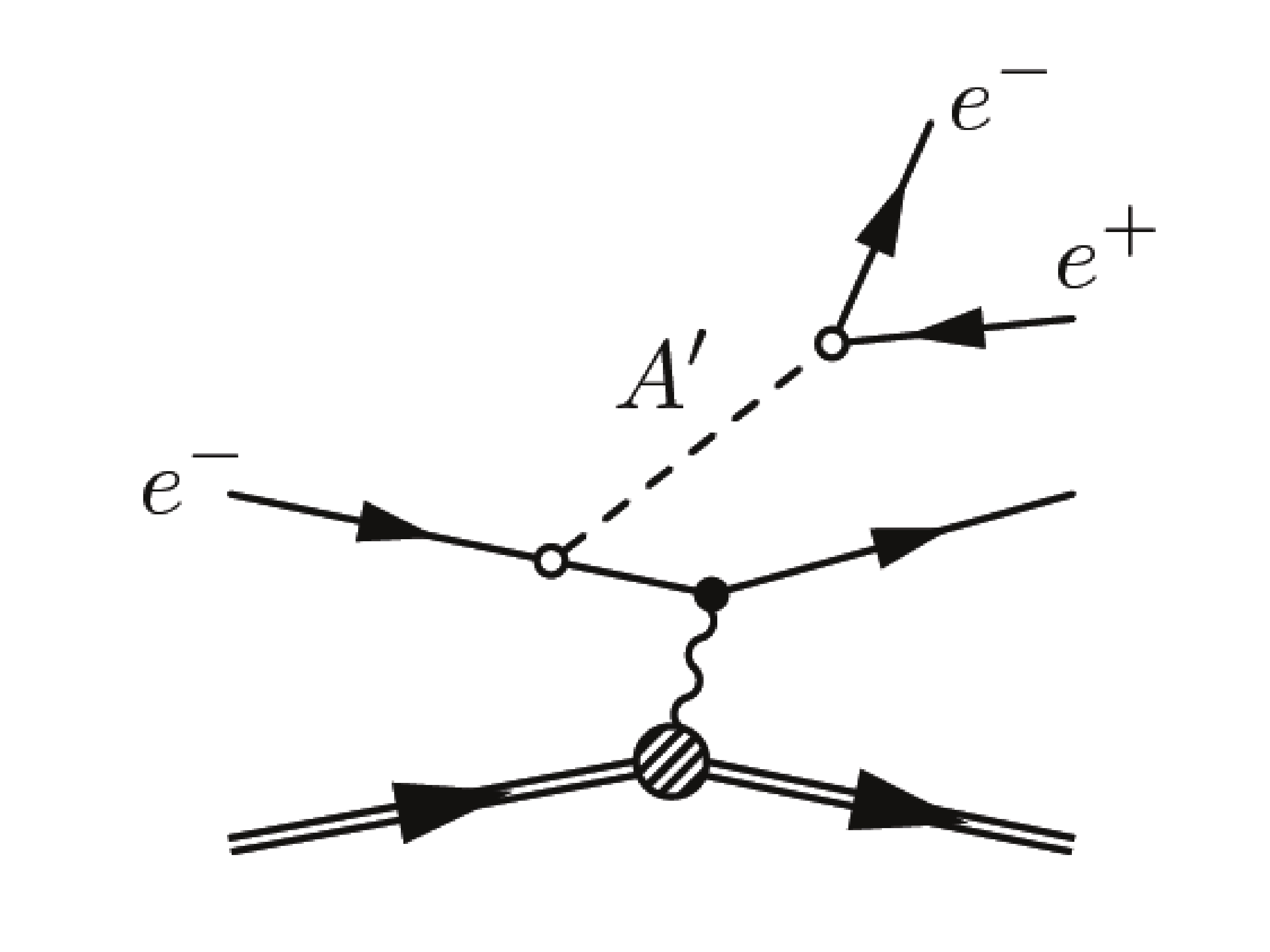}
\caption{\label{fig:Aprod} Reaction mechanism for $A^\prime$ production via bremsstrahlung in the interaction of an electron with a nuclear target.}
\end{figure}

\begin{figure}
\includegraphics[width=4. in]{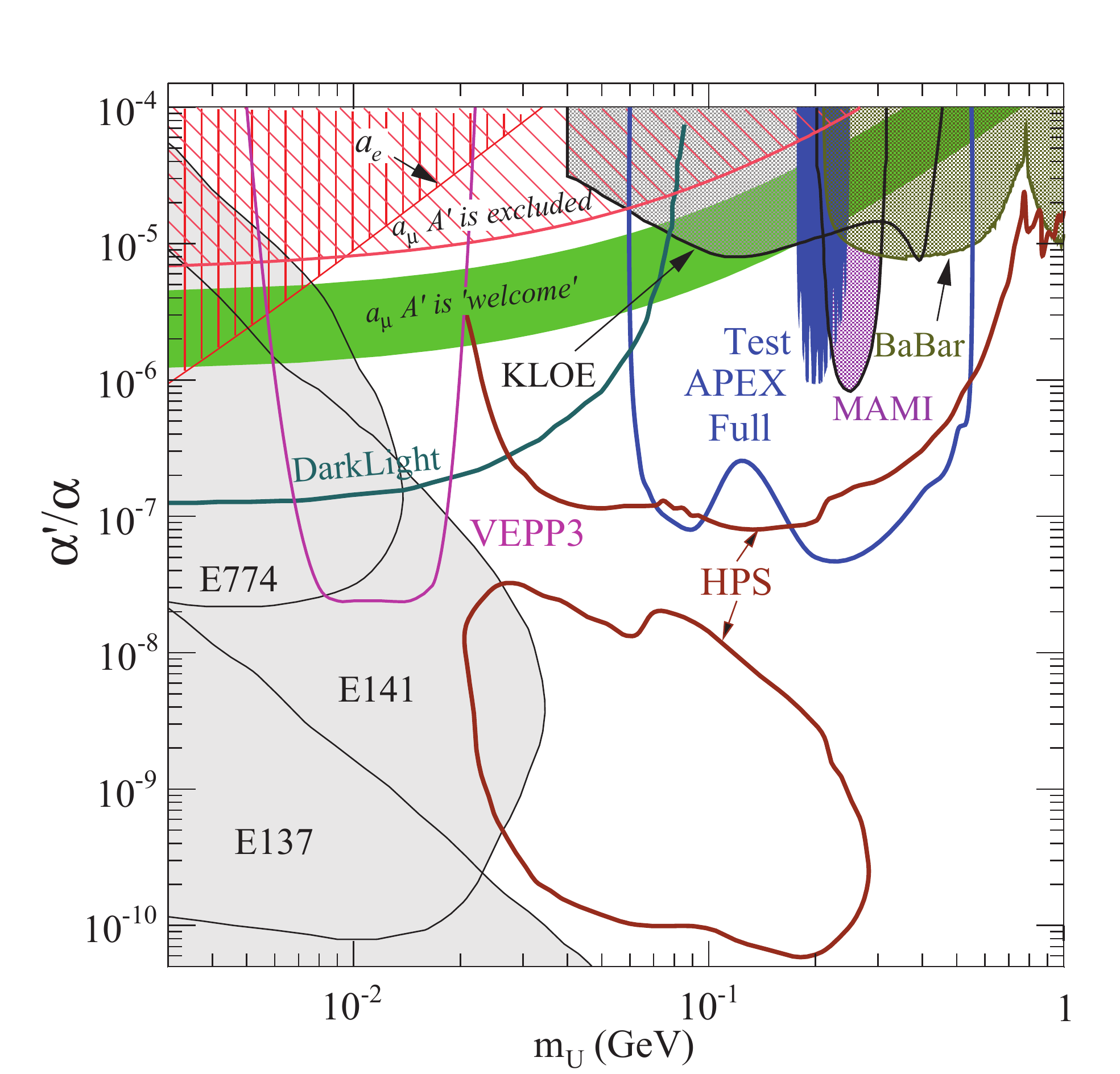}
\caption{\label{fig:Aprime} Projected sensitivities ($2\sigma$) of Jefferson Lab $A^\prime$ proposals along with limits (in gray) from previous experiments, shown as a function of the $A^\prime$ mass and the interaction strength $\alpha^\prime/\alpha = \epsilon^2$. The green band indicates the prediction resulting from interpretation of the muonic $g-2$ anomaly as due to an $A^\prime$. }
\end{figure}

Recently, it has been postulated that WIMPs or other dark matter particles could interact with normal matter through the exchange of a new vector gauge boson \cite{Ap}. Such a vector boson would be a very natural ingredient in theories that include WIMPs \cite{Holdom}. In addition, such particles could help explain other astrophysical anomalies associated with excess positrons and/or electrons at multi-GeV energies. Such a new vector boson would couple to electrons, with a coupling constant $\epsilon$. Moreover, it has been realized \cite{Bj} that the present experimental limits in the mass range 10-1000~MeV do not strongly constrain the existence of such a particle with electron couplings up to about $\epsilon \sim $~few~$\times 10^{-3}$. This realization has motivated several new proposals to Jefferson Lab for experiments to search for such a new neutral weak vector boson, now referred to as $A^\prime$. In these experiments the $A^\prime$ is produced via a bremsstrahlung process (Fig.~\ref{fig:Aprod}) and decays to an $e^+-e^-$ pair. The projected sensitivities for these proposed experiments is displayed in Fig.~\ref{fig:Aprime}.

\subsection{APEX in Hall A}
The APEX collaboration has proposed \cite{APEX} to use the CEBAF electron beam at energies of 1-4 GeV incident on 0.5-10\%
radiation length Tungsten ribbon targets.  The $e^+e^-$ pairs at small angles, $\theta \simeq 5^\circ$, from the produced $A^\prime$ decay are detected using the High Resolution Spectrometers and the septum magnet in Hall A. The experiment will scan a range of $A^\prime$ masses for each kinematic setting to cover the range 60-1000~MeV. This experiment utilizes primarily existing equipment in Hall A. The collaboration has recently completed a successful test run in Hall A demonstrating the feasibility of this approach \cite{APEXtest}.

\subsection{HPS - Heavy Photon Search in Hall B}

The Heavy Photon Search (HPS) proposal \cite{HPS} would site an experimental apparatus in Hall B, downstream of the CLAS12 detector. This experiment would utilize a low current electron beam ($<1~\mu$A) at 4-11~GeV on a tungsten target (0.14-1\% r.l.). Downstream of the target will be a dipole analyzing magnet (0.917~T-m), followed by silicon trackers (12 planes over 1~m length), an electromagnetic calorimeter, and a muon detector (steel plates with scintillation hodoscopes). The dipole magnet will spread the high flux of radiation from the target into a ``sheet of flame'', and the detectors must be split to transmit these particles and still operate in a high rate environment. The unique features of this experiment include the capability to detect pairs from decays downstream of the target (longer lifetime $A^\prime$ corresponding to lower values of $\epsilon$) and the ability to also detect muon pairs.

\subsection{DARKLIGHT at the FEL}

The DARKLIGHT experiment \cite{DARKLIGHT} is a proposal to extend the search for $A^\prime$ to lower mass values, down to $\sim 10$~MeV. This experiment would utilize the high intensity (10~mA) electron beam at 140~MeV available at the Free Electron Laser (FEL) facility at Jefferson Lab, incident on a $10^{19}$~cm$^{-2}$ gas hydrogen target. A magnetic spectrometer detects all three leptons and a high resolution detector a few centimeters from the interaction region detects the final state protons.  Measurement of all four final state particles and good momentum resolution allows reconstruction of the A' mass with 1~MeV precision over the mass range 10-100~MeV.

\section{Summary}

The unique facilities available at Jefferson Lab will
provide many opportunities for exploration and discovery for a large
international community of nuclear scientists for many years to come.
In particular, the remarkable and unique electron beam quality at Jefferson Lab enables a powerful
new experimental program that offers opportunities for discovery of new
physics beyond the standard model. Previous experience with parity-violation
experiments provides a foundation for higher precision measurements with a physics
reach comparable or exceeding the Large Hadron Collider. The development of detector systems
capable of high luminosity operation with open geometry at Jefferson Lab now enables
well-motivated searches for new vector gauge bosons that are favored in many dark matter scenarios.
This new program effectively utilizes the existing infrastructure at the
Laboratory but also requires the construction of new experimental apparatus.

\begin{theacknowledgments}
 I am grateful to Roger Carlini, Krishna Kumar, Paul Souder, Jens Erler, Peter Fisher, Stepan Stepanyan, and Bogdan  Wojtsekhowski
for their assistance in preparing this document. This work was
supported by DOE contract DE-AC05-06OR23177, under which Jefferson
Science Associates, LLC, operates the Thomas Jefferson National
Accelerator Facility.
\end{theacknowledgments}


\bibliographystyle{aipproc}   

\bibliography{sample}


\end{document}